\documentclass[a4paper,11pt]{article}
\pdfoutput=1 

\usepackage{jheppub} 

\usepackage[T1]{fontenc} 

\usepackage{graphicx}
\usepackage{comment}

\def\no{\nonumber}

\newcommand{\p}{\partial}

\newcommand{\tr}{\textrm{Tr}}

\title{\boldmath Finite temperature holographic duals of $2$-dimensional BCFTs}

\author[a]{J. Estes,}

\affiliation[a]{Blackett Laboratory, Imperial College,\\London, SW7 2AZ, United Kingdom}

\emailAdd{j.estes@imperial.ac.uk}

\abstract{We consider holographic duals of $2$-dimensional conformal field theories in the presence of a boundary, interface, defect and/or junction, referred to collectively as BCFTs. In general, the presence of a boundary reduces the $SO(2,2)$ conformal symmetry to $SO(2,1)$ and the dual geometry is realized as a warped product of the form $AdS_2 \times {\cal M}$, where ${\cal M}$ is not compact.  In particular, it will contain points where the warp factor of the $AdS_2$ space diverges, leading to asymptotically $AdS_3$ regions.  We show that the $AdS_2$ space-time may always be replaced with an $AdS_2$-``black-hole'' space-time.  We argue the resulting geometry describes the BCFT at finite temperature.  To motivate this claim, we compute the entanglement entropy holographically for a segment centered around the defect or ending on the boundary and find agreement with a known universal formula.}

\preprint{Imperial/TP/2015/JE/01}

\begin{document}

\maketitle
\flushbottom

\section{Introduction}
\label{sec:intro}

The AdS/CFT correspondence allows one to study strongly coupled systems by instead studying weakly coupled dual gravitational systems.  In general, the isometry of the supergravity solution matches the conformal symmetry of the dual field theory.  For a $d$-dimensional conformal field theory with symmetry $SO(2,d)$, this fixes the geometry to be of the form $AdS_{d+1} \times {\cal M}$ with ${\cal M}$ a compact manifold.  The class of holographic conformal fixed points can be enlarged by allowing for the presence of conformal interfaces, defects, boundaries or junctions (BCFTs).  For $2$-dimensional theories, BCFTs can be constructed in string theory using the D1/D5 system \cite{Chiodaroli:2009yw,Chiodaroli:2009xh,Chiodaroli:2011nr} and can be thought of as a junction of quantum critical wires.  Defects, corresponding to D3-branes, can be inserted at the junction \cite{Chiodaroli:2012vc} and one may consider a single wire ending on a defect, leading to a boundary conformal field theory \cite{Chiodaroli:2011fn}.

The presence of an interface, defect or boundary can lead to interesting phenomena in the field theory.  For example, a magnetic impurity can lead to the Kondo effect \cite{Kondo01071964}, a boundary may exhibit surface superconductivity \cite{PhysRevLett.12.442} and interesting edge states arise at the interface of two systems in different topological phases \cite{PhysRevLett.49.405,2009AIPC.1134...22K,RevModPhys.82.3045}.  In order to study such phases of matter in the context of AdS/CFT, it is necessary to go beyond the conformal field theory description.  In practice, this means studying the system in a state which is not simply empty vacuum.  The two main ingredients one introduces are temperature and chemical potentials for various global symmetries.  This is done by introducing a charged black hole or domain wall into the $AdS_{d+1} \times {\cal M}$ space-time.  The phase structure of the field theory system is then mapped onto the phase structure of black hole and domain wall solutions with $AdS_{d+1} \times {\cal M}$ asymptotics.

\begin{figure}[h!]
  \centering
    \includegraphics[width=0.7\textwidth]{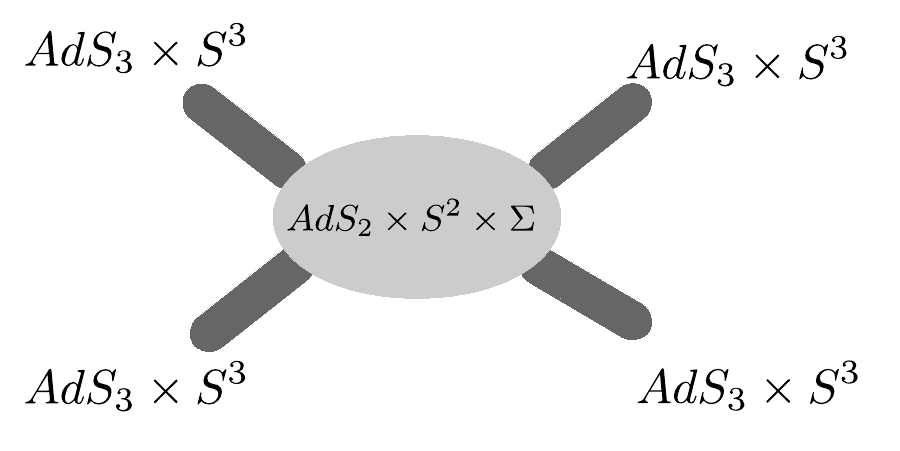}
    \caption{Geometry from \cite{Chiodaroli:2011nr} corresponding to a 4-junction of quantum wires.  Each leg (dark region), corresponds to an asymptotically $AdS_3 \times S^3$ region and is covered by a Fefferman-Graham coordinate system.  The center (light region), corresponds to a space of the form $AdS_2 \times S^2 \times \Sigma$ with $\Sigma$ a $2$-dimensional compact space.  Note that this space is not covered by asymptotically $AdS_3$ Fefferman-Graham coordinates.}
    \label{fig1}
\end{figure}

When one has the full conformal isometry group, one can often work with a truncated theory in the lower dimensional $AdS_{d+1}$ space-time.  The problem then reduces to the simpler problem of studying charged black holes and domain walls in asymptotically $AdS_{d+1}$ space-time.  However, in the presence of a defect, interface or boundary, the problem becomes much more difficult and consistent truncations are difficult to find.  In most examples the geometry cannot be covered by a single set of Fefferman-Graham coordinates, since the boundary is now only locally asymptotically $AdS_{d+1}$.  An example is depicted in figure \ref{fig1}.

In \cite{Bak:2011ga}, the finite temperature versions of Janus solutions were constructed.  In this paper we generalize their results by giving a prescription for inserting neutral black holes into arbitrary space-times which are dual to $2$-dimensional boundary/interface/defect/junction conformal field theories.
To simplify the discussion, we shall refer to all cases as simply boundary conformal field theories.\footnote{Using the folding trick, all cases can be related to boundary conformal field theories.}
The presence of the boundary reduces the conformal symmetry from $SO(2,2)$ to $SO(2,1)$.  For such theories the geometries corresponding to the conformal point always take the form $AdS_{2} \times \hat {\cal M}$, where $\hat {\cal M}$ is not compact.  In particular, it will contain points where it combines with the $AdS_2$ fiber to form asymptotically $AdS_3 \times {\cal M}$ geometries.

In \cite{Bachas:2011xa}, it was argued that at the linear level, we may always replace the $AdS_{2}$ fiber with any space whose metric is Einstein, or equivalently with any space which satisfies the same linearized Einstein equations as the $AdS_2$ space.  Here we argue that this extends to the non-linear level. In particular we may replace $AdS_2$ with a $2$-dimensional version of the $AdS$-Schwarzschild solution.  We show that for $AdS_3$, this procedure correctly generates the BTZ black-hole.  We then give further evidence that this corresponds to putting the field theory at finite temperature by computing the entanglement entropy corresponding to a segment centered around the interface/defect or ending on the boundary.  Our holographic results agree with the general result found for $2$-dimensional conformal field theories given in \cite{Calabrese:2004eu,Calabrese:2009qy}.  In particular, for a junction of $N$ quantum critical wires, we show that the holographic entanglement entropy takes the form
\begin{align}
S_{N} =& \sum_{i=1}^N \left[ \frac{c_i}{6} \ln \left( \frac{\beta}{\pi a} \sinh \frac{2\pi\ell}{\beta} \right)\right] + \tilde c_1 \,,
\end{align}
where $\beta^{-1}$ is the temperature and the central charge of the theory living on the $i$-th wire is given by $c_i$.  The quantity $\tilde c_1$ is a boundary/defect entropy.  In particular, the contribution of the boundary/defect to the entanglement entropy is not modified by thermal effects.
Finally, we note that for the $AdS_{3}$ Janus black hole solution of \cite{Bak:2011ga}, their analytic solution given in $(5.6)$ of \cite{Bak:2011ga} follows from the above procedure.  Their final metric is given by a warped product of the form $ds_3^2 = f(\mu)(ds^2_2 + d\mu^2)$, where $ds^2_2$ is given the metric on $AdS$-Schwarzschild and the warp factor $f(\mu)$ is determined by the original zero-temperature Janus solution (see the discussion above (3.22) of \cite{Bak:2011ga}).

We note that for the higher dimensional cases, we will again have $AdS$ fibers which may be replaced with corresponding $AdS$-Schwarzschild solutions. This will always yield a solution to the equations of motion and so one may wonder if this corresponds to putting the theory at finite temperature. Upon inspecting the asymptotics of the modified solution, one will find that the boundary metric is not conformally flat.  Namely, this modification corresponds not only to putting the theory at finite temperature, but also to putting the field theory on a curved space-time.  In the special case of $2$-dimensions, all Riemannian spaces are (locally) conformally flat.  Thus only for $2$-dimensions were we guaranteed to find a coordinate system in which the boundary metric was flat.\footnote{In general, the metric is only locally conformally flat.  This is related to the fact that we cannot cover the manifold with a single set of Fefferman-Graham coordinates.  However, in each Fefferman-Graham patch, we were guaranteed to find appropriate coordinates in which the induced metric on the boundary was flat.}  We propose that this fact is related to the lack of temperature dependance appearing in the boundary entropy for $2$-dimensional conformal field theories and that for higher dimensions, the corresponding boundary entropy will in general depend on the temperature.  In particular, one may form dimensionless quantities from combinations involving the temperature and invariants which characterize the entangling surface.

The remainder of the paper is organized as follows.  In section \ref{sec:2}, we explicitly show how to construct a finite temperature gravity solution from a given $AdS_2$-sliced domain wall solution supported by a single scalar field, corresponding to a $2$-dimensional interface CFT.  We then show how our finite temperature prescription generalizes to arbitrary gravitational systems with solutions which describe $2$-dimensional boundary, interface or junction CFTs.  In section \ref{sec:3}, we compute holographically the general form of the entanglement entropy for a single line segment centered around the defect and show agreement with the known result from the field theory side.  In particular, we obtain an exact match for the temperature dependence between the field theory and gravitational descriptions.  In appendix \ref{app:sixdcase}, we give the explicit finite temperature generalization of the solutions constructed in \cite{Chiodaroli:2011nr}, which describe supersymmetric junctions of $(p,q)$-strings.

\section{Black hole solutions}
\label{sec:2}

We consider first the simple case of a $3$-dimensional Janus solution \cite{Bak:2003jk} or a more general Janus type solution supported by a scalar field with non-trivial potential, as in \cite{Korovin:2013gha}.  Such solutions can be thought of as $AdS_2$-sliced domain walls.  The geometry preserves an $SO(2,1)$ symmetry and is locally asymptotically $AdS_3$.  By locally, we mean that the $AdS_3$ boundary is split into two halves and the scalar field may take different values on the two halves.  Additionally, the scalar and metric field may be sourced along the intersection of the two boundary components.  In the context of AdS/CFT, such geometries are dual to conformal theories with interfaces and defects.

In this simple case, the field content consists of gravity coupled to a scalar field and a cosmological constant as in \cite{Korovin:2013gha}.  We will discuss more general theories below.  We take the action to be given by
\begin{align}
S &= - \frac{1}{2 \kappa^2} \int d^{3}x \sqrt{|g|} \left( R + \frac{2}{L^2} - \frac{1}{2} (\p \phi)^2 - V(\phi) \right) \,.
\end{align}
The corresponding equations of motion are given by
\begin{align}
&R_{\mu \nu} + \frac{2}{L^2} g_{\mu \nu} =  \frac{1}{2} \p_\mu \phi \p_\nu \phi  - \frac{1}{2} g_{\mu \nu} V(\phi)\,, \cr
&\frac{1}{\sqrt{|g|}} \p_\nu\left( \sqrt{|g|} g^{\mu \nu} \p_\mu \phi \right) - \frac{\p V}{\p \phi} = 0\,.
\end{align}
We assume that the scalar potential has at least one minimum $\phi_0$ with $V(\phi_0) = 0$.  Note that we can always set $V(\phi_0) = 0$ by a redefinition of $L$.  In this case, $AdS_3$ with radius $L$ and $\phi = \phi_0$ is a solution to the equations of motion.  In general, one may choose the potential $V$ such that the equations of motion admit $AdS_2$-sliced domain wall solutions.  Such solutions have been classified in \cite{Korovin:2013gha} and several families have been analytically constructed.  Note that when $V$ has two minima, $\phi_\pm$, one has the possibility to construct solutions where the $AdS_2$-sliced domain wall interpolates from $\phi_-$ to $\phi_+$ and the asymptotic $AdS_3$ radius can take different values at the two endpoints of the interpolation, with the value of the radius $L_\pm$ determined by the value of the potential $V(\phi_\pm)$.

In general the Janus type solutions take the form
\begin{align}
\label{JanusAns}
&ds_{CFT}^2 = f(x)^2 ds^2_{AdS_2} + dx^2\,,&
&\phi = \phi(x)\,,&
\end{align}
where $ds^2_{AdS_2}$ is the metric on $AdS_2$ with unit radius.  In these coordinates, one can easily see the $AdS_2$-domain wall structure.
We introduce the frames $e^a = dx$ and $e^m = f(x) \hat e^m$ with $m=1,2$.  The $\hat e^m$ are frames on the unit $AdS_2$.
Given the ansatz \eqref{JanusAns}, the equations of motion reduce to
\begin{align}
\label{CFTeom}
&R_{mn} = \left(-\frac{1}{f^2} - (\p_x \ln f)^2  - \frac{\p_x^2 f}{f} \right) \eta_{mn}
 = - \frac{2}{L^2} \eta_{mn} - \frac{1}{2} \eta_{mn} V(\phi)\,, \cr
&R_{aa} = - 2 \frac{\p_x^2 f}{f} = - \frac{2}{L^2} + \frac{1}{2} (\p_x \phi)^2 - \frac{1}{2} V(\phi)\,, \cr
&\p_x^2 \phi + 2 (\p_x \ln f) (\p_x \phi) = \frac{\p V}{\p \phi}\,.
\end{align}
We assume that there exists an $f$ and $\phi$ which satisfy these equations and that the geometry is (locally) asymptotically $AdS_3$.  Such $f$ and $\phi$ can be obtained from the formalism developed in \cite{Korovin:2013gha}.  There one picks a pair $(f,\phi)$, subject to certain constraints, and then constructs the appropriate potential $V$.

Since the geometry is asymptotically $AdS_3$, in the asymptotic regions, there will exist a map to Fefferman-Graham coordinates \cite{MR837196} (see also \cite{deHaro:2000xn}).  In these coordinates, the metric takes the general form
\begin{align}
ds^2 = L^2 \left( \frac{dr^2}{r^2} + \frac{g_{\alpha \beta}}{r^2} dy^{\alpha} dy^{\beta} \right)\,,
\end{align}
with $g_{\alpha \beta} = g^{(0)}_{\alpha \beta} + g^{(2)}_{\alpha \beta} r^2 + ...$\,.  Using the scaling symmetry and $1+1$-dimensional Poincar\'e symmetry, the metric is restricted to take the form
\begin{align}
ds^2 = L^2 \left( \frac{dr^2}{r^2} + \frac{g_{yy}}{r^2} dy^2 + \frac{g_{tt}}{r^2} dt^2 \right)
\end{align}
where $g_{yy}$ and $g_{tt}$ are functions of $r/y$.  To exhibit the explicit map, we first introduce coordinates for $AdS_2$ with $ds^2_{AdS_2} = \rho^{-2} d\rho^2 - \rho^2 dt^2$.  The explicit coordinate transformation is then given by
\begin{align}
\label{CFTFGtrans}
&r = \frac{\exp\left(-\int d\tilde x \, \frac{1}{f(\tilde x)} \sqrt{\frac{f^2(\tilde x)}{L^2} - 1}\right)}{\rho}\,,&
&y= \pm \frac{\exp\left(\int d\tilde x \, \frac{1}{f(\tilde x) \sqrt{\frac{f^2(\tilde x)}{L^2} - 1}}\right)}{\rho}\,,&
\end{align}
where the the overall sign choice for the root is determined by the requirement that $r \rightarrow 0$ as $x \rightarrow \pm \infty$.  The metric factors are given by
\begin{align}
&g_{yy} = \frac{r^2}{y^2} \left( \frac{f^2}{L^2} - 1\right)\,,&
&g_{tt} = - r^2 \rho^2 \frac{f^2}{L^2} \,.&
\end{align}
The integration constants in \eqref{CFTFGtrans} can be fixed by requiring $g_{yy}(0) = - g_{tt}(0) = 1$.  We note here that the Fefferman-Graham coordinates typically do not cover the entire space \cite{Estes:2014hka}.  Indeed, it can be observed that they break down whenever $f < L$.

For the case of $AdS_3$ itself, we have $f = L \cosh(x/L)$ and
\begin{align}
&r = \frac{1}{\cosh(x/L) \rho}\,,&
&y = \frac{\tanh(x/L)}{\rho}\,,&
\end{align}
along with $g_{yy} = 1$ and $g_{tt} = -1$ as expected.  For the more general case, where space-times is only asymptotically $AdS_3$, we require the warp factor, $f(x)$, to have the same leading behavior as the $AdS_3$ case.  Namely, as $x \rightarrow \pm \infty$, we require $f \sim f^\pm_{-1} e^{\pm x/L} + ...$ where the additional terms vanish in the $x \rightarrow \pm \infty$ limit.  Note that we have allowed for an arbitrary shift in $x$ by introducing the overall constant $f_{-1}^\pm$.  Making use of \eqref{CFTFGtrans}, we find at leading order
\begin{align}
x =& \mp \left[L \ln(r) - \frac{L}{2} \ln \left( \frac{L^4 y^2}{f_{-1}^2} \right) + \frac{r^2}{4 L y^2} + {\cal O}(r^4)\right]\,, \cr
\rho =& \pm \left[\frac{1}{y} - \frac{r^2}{2 L^2 y^3} + {\cal O}(r^4)\right]\,,
\end{align}
along with $g_{yy} = 1 + {\cal O}(r^2)$ and $g_{tt} = -1 + {\cal O}(r^2)$ where the additional terms vanish in the $x \rightarrow \pm \infty$ limit.  Thus we see that the boundary metric is flat.

We now consider the more general anstaz
\begin{align}
\label{metgen}
&ds^2 = f(x)^2 ds^2_{1,1} + dx^2\,,&
&\phi = \phi(x)\,,&
\end{align}
where $ds^2_{1,1}$ is an arbitrary metric on the two-dimensional space with signature $(-1,+1)$ and curvature $\hat R_{mn}$.  In particular, we take $ds^2_{1,1}$ and correspondingly $\hat R_{mn}$ to be independent of $x$.  In this case, the equations of motion reduce to
\begin{align}
\label{genEOM}
&R_{mn} = \left(\frac{\hat R_{mn}}{f^2} - (\p_x \ln f)^2  - \frac{\p_x^2 f}{f} \right) \eta_{mn}
 = - \frac{2}{L^2} \eta_{mn} - \frac{1}{2} \eta_{mn} V(\phi) \cr
&R_{aa} = - 2 \frac{\p_x^2 f}{f} = - \frac{2}{L^2} + \frac{1}{2} (\p_x \phi)^2 - \frac{1}{2} V(\phi) \cr
&\p_x^2 \phi + 2 (\p_x \ln f) (\p_x \phi) = \frac{\p V}{\p \phi}
\end{align}
If we require the metric on $ds^2_{1,1}$ to again be Einstein so that
\begin{align}
\label{redEOM}
\hat R_{mn} = - \eta_{mn} \,
\end{align}
then the above equations reduce to \eqref{CFTeom}.  Thus we again obtain a solution to the equations of motion, where $f(x)$ is the same function as in the conformal case.

If we require the metric to be static (independent of $t$), then the general solution to \eqref{redEOM} takes the form
\begin{align}
\label{Tfib}
ds^2_{1,1} = - (\rho^2-1)dt^2 + \frac{d\rho^2}{\rho^2-1} \, ,
\end{align}
with $1 < \rho < \infty$.
It is non-trivial to obtain a set of Fefferman-Graham coordinates with flat boundary metric.\footnote{A simple generalization of \eqref{CFTFGtrans} is given by taking
\begin{align}
&r = \frac{\exp\left(-\int d\tilde x \, \frac{1}{f(\tilde x)} \sqrt{\frac{f^2(\tilde x)}{L^2} - 1}\right)}{\rho + \sqrt{\rho^2-1}}\,,&
&y= \frac{\exp\left(\int d\tilde x \, \frac{1}{f(\tilde x) \sqrt{\frac{f^2(\tilde x)}{L^2} - 1}}\right)}{\rho + \sqrt{\rho^2-1}}\,.&
\end{align}
However, working out the asymptotics one finds that the boundary metric is not flat in these coordinates.  In particular, $g_{tt}$ is not constant at $r=0$ (on the boundary), but depends on $y$.
}
To obtain a set of Fefferman-Graham coordinates with flat boundary metric, we work in an expansion as $x \rightarrow \pm \infty$.  Typically $f$ will have the following form as an asymptotic expansion\footnote{We note that allowing polynomial $x$-dependence in the coefficients, yields logarithmic terms in the Fefferman-Graham expansion.}
\begin{align}
f = f^{\pm}_{-1} e^{\pm x/L}  + \sum_{n=1}^\infty f^{\pm}_n e^{\mp n x / L} \,.
\end{align}
The transformation to Fefferman-Graham coordinates with flat boundary metric can be worked out order by order in $r$.  The expansion takes the form
\begin{align}
\label{flatFG}
&x = \mp L \ln(r) \pm \frac{L}{2} \ln \left( \frac{L \sinh(y)}{f^\pm_{-1}} \right)^2 \pm \frac{L}{4} \coth^2(y) r^2 + {\cal O}(r^3)\,,\cr
&\rho = \pm \coth(y) \pm \frac{1}{2} \frac{\cosh(y)}{\sinh^3(y)} r^2 + {\cal O}(r^3)\,,
\end{align}
where the upper sign is valid for $y>0$ and the lower sign is valid for $y<0$.  At leading order the metric is given by
\begin{align}
\label{flatFGmet}
g_{yy} =& 1 + \frac{1}{2} \left[1 - \frac{1}{\sinh^2(y)}\left(1 - 4 \frac{f^\pm_{-1} f^\pm_1}{L^2} \right) \right]r^2  + {\cal O}(r^3) \,,\cr
g_{tt} =& -1 + \frac{1}{2} \left(\coth^2(y) - 4 \frac{f^\pm_{-1} f^\pm_1}{\sinh^2(y) L^2}  \right)r^2 + {\cal O}(r^3) \,.
\end{align}
Thus the boundary metric is flat as desired.  The remaining coefficients can be computed recursively by solving algebraic equations.  We note, as in the conformal case, that these expansions generically do not cover the entire space.  In particular, the higher order terms are suppressed only when $r \ll y$.  That is, the expansions are valid as we approach the boundary away from the defect/interface.

Taking $f = L \cosh(x/L)$, we should recover the BTZ black hole.  To see this, we start with the BTZ black hole metric
\begin{align}
\label{BTZ}
ds^2_{BTZ} = -\frac{z^2-z_0^2}{L^2} dt'^2 + \frac{L^2}{z^2-z_0^2} dz^2 + z^2 dy'^2 \,.
\end{align}
This corresponds to a BTZ black hole with Hawking temperature $T = z_0/2 \pi L^2$.\footnote{To see this, note that in Euclidean signature, the metric is regular provided $\tau = i t'$ has periodicity $2 \pi L^2/z_0$.}  Next we change to Fefferman-Graham coordinates by introducing $r$ as $z = (4+r^2)z_0/4r$.  We also scale $t'$ and $y'$ as $t = z_0 t'/L^2$ and $y = z_0 y'/L$.  In these coordinates, the metric is given by
\begin{align}
ds^2_{BTZ} = L^2 \left(\frac{dr^2}{r^2} - \frac{(r^2-4)^2}{16r^2} dt^2 + \frac{(r^2 + 4)^2}{16 r^2} dy^2 \right)\,.
\end{align}

Next, we compute the Fefferman-Graham expansion for the metric \eqref{metgen} taking $f = L \cosh(x/L)$.  We make use of \eqref{flatFG} and \eqref{flatFGmet} and have computed the expansions explicitly to tenth order in $r$.  For the metric factors, we find
\begin{align}
&g_{yy} = 1 + \frac{r^2}{2} + \frac{r^4}{16} + {\cal O}(r^{11}) \,,&
&g_{tt} = -1 + \frac{r^2}{2} - \frac{r^4}{16} + {\cal O}(r^{11}) \,.&
\end{align}
Thus the metric factors exactly reproduce the BTZ black hole up to the order we have computed at.  Since we have matched the leading and subleading terms of the Fefferman-Graham expansion, we conclude, using the recursion developed in \cite{deHaro:2000xn}, that we have exactly the BTZ black hole.  The coordinate transformation giving $x$ and $\rho$ in \eqref{flatFG} is complicated and we do not give the expressions here.  We note that the expansion does not terminate at any finite order in $r$, unlike the expansions for the metric factors.

To summarize, we have shown that given a geometry dual to a $2$-dimensional defect/interface CFT, we may always find a new solution to the equations of motion by replacing the $AdS_2$ fiber with the space given in \eqref{Tfib}.  There always exists a choice of Fefferman-Graham coordinates such that the metric induced on the boundary is flat.  Finally, we can associate a temperature to the solution by requiring the Euclidean metric to be regular, which implies that $\tau = i t$ has periodicity $2 \pi$ and the corresponding temperature is $T = 1/2 \pi$.  To obtain more general temperatures, we may introduce a scaled time variable $t'$ as in the BTZ black hole case above, so that $t = z_0 t'/L^2$ and the temperature is $T = z_0/2 \pi L^2$.  We shall make this more precise by computing the entanglement entropy in the next section.

Although we have only considered the simple system consisting of gravity, a cosmological constant and a scalar field, our analysis extends to more general theories.  First, we may introduce additional scalar fields without modifying the above analysis.  Secondly, we may consider higher dimensional theories of gravity.  In general, a geometry which describes a $2$-dimensional defect/interface CFT will take the form
\begin{align}
ds_{CFT}^2 = f(y_a)^2 ds^2_{AdS_2} + ds^2_{y_a}\,.
\end{align}
We first assume that a solution to the equations of motion exists, then we observe that we can find a new solution by replacing $ds^2_{AdS_2}$ with \eqref{Tfib}.  To see this, we note that the Ricci tensor again takes the same form as given in \eqref{genEOM}, except that the expression for $R_{ab}$ will be more complicated.  In particular, after replacing $ds^2_{AdS_2}$ with an arbitrary $2$-dimensional metric, the only dependence on this metric appears as $\hat R_{mn}$ in the first term of $R_{mn}$ given in \eqref{genEOM}.

So far, we have discussed only the case of interface conformal field theories.  For a boundary conformal field theory, there will only exist a single point in $ds^2_{y_a}$ where the warp factor $f$ diverges.  We can always pick local polar coordinates, $\{x, w^{\tilde a}\}$, around the point, where $x^{-1}$ is a radial coordinate, the $w^{\tilde a}$ are the angular coordinates and $f$ behaves as $f \sim f_{-1} e^{x/L}$, as $x \rightarrow \infty$.  Here, we only have a single Fefferman-Graham patch and the boundary geometry is a half space.  In this case, the boundary of the CFT can either give rise to an explicit boundary in the dual geometry, such as in \cite{Takayanagi:2011zk,Erdmenger:2014xya}, or it can be realized without an explicit boundary by making use of the internal geometry as in \cite{Chiodaroli:2012vc,Chiodaroli:2011fn}.  In this paper, we focus on the second case, while additional subtleties arise in the first case.\footnote{From the lower dimensional point of view, the second case may be thought of as the first case with a particular choice of boundary conditions arising from the dimensional reduction.}

For an $N$-junction, we will have $N$ points where the warp factor diverges.  At the $i$-th point we can again pick local polar coordinates $\{x, w^{\tilde a}\}$  with $f \sim f^{i}_{-1} e^{x/L_{i}}$, as $x \rightarrow \infty$.  In this case, we have $N$ Fefferman-Graham patches and the boundary geometry is $N$ half spaces, which are all glued together at the boundary of the $AdS_2$ fiber.  For the special case of $N=2$, the boundary simply becomes the full plane after gluing.  In all cases the above black-hole construction proceeds as before.  We simply replace the $AdS_2$ fiber with the metric \eqref{Tfib}.

Finally, we also consider the case of form fields.  Generally, the conformal symmetry dictates that any form field either be proportional to the volume form of $AdS_2$ or have no support along $AdS_2$.  In the latter case, modifying the $AdS_2$ metric has no effect on the form field.  In the former case, we first write the form field as $F = F_{mn} \hat e^{mn} \wedge F_y$ where $\hat e^{mn}$ is the unit volume form on $AdS_2$ and $F_y$ are the components of $F$ along $ds^2_{y_a}$.  We now replace $ds^2_{AdS_2}$ with an arbitrary $2$-dimensional metric, while holding fixed the component $F_{mn}$.  With this prescription, the contribution of $F$ to the stress energy tensor, $T_{mn}$ and $T_{ab}$ does not change.  Thus we will find that taking $ds^2_{1,1}$ to be given by \eqref{Tfib} satisfies the Einstein equations.  Typically the equation of motion for $F$ itself takes the form $* d * F = 0$.  Since $F$ is still proportional to the $2$-dimensional volume form, its exterior derivative along the $2$-dimensional space will still be zero and the equation of motion will again automatically be satisfied.  In appendix \ref{app:sixdcase}, we illustrate this discussion with an explicit example from six dimensional supergravity.   In particular, we consider the solutions of \cite{Chiodaroli:2011nr}, which describe junctions of $(p,q)$-strings, as well as the more general solutions of \cite{Chiodaroli:2012vc,Chiodaroli:2011fn} which allow for boundaries and defects.

\section{Entanglement entropy}
\label{sec:3}

A straightforward way to exhibit the thermal nature of the solutions of section \ref{sec:2} would be to compute the free energy.  However, a precise computation requires the introduction of a cut-off surface followed by holographic regularization.  This process is complicated in our geometries, as the geometry cannot be covered by a single Fefferman-Graham patch.

As an alternative to exhibit the thermal nature of our solution, we compute the entanglement entropy of a single line segment.  To define entanglement entropy one first divides the system into two subsystems, which we call $A$ and $B$. Here, we take $A$ to be a segment of length $\ell$ and $B$ to be its complement.  The total Hilbert space $H$ is then given by the product $H = H_A \otimes H_B$.  We define a reduced density matrix by first tracing over all states in $B$, $\rho_A = \tr_{H_B} \rho$, where $\rho$ is the density matrix.  For a pure state, $\rho$ is simply the projection matrix onto the ground state, while for a thermal system we have $\rho = \exp(-\beta H)$.  The entanglement entropy is defined in terms of $\rho_A$ by
\begin{align}
S = - \tr_{H_A} \rho_A \ln \rho_A \,.
\end{align}
In \cite{Holzhey:1994we}, it was shown that for a $2$-dimensional conformal field theory, the entanglement entropy for a single interval of length $\ell$ takes the form
\begin{align}
\label{entCFT}
S = \frac{c}{3} \ln \left(\frac{\ell}{a}\right) + \hat c_1 \,,
\end{align}
where $c$ is the central charge, $a$ is a UV-cutoff (or lattice spacing) and $\hat c_1$ is a non-universal constant.  In the presence of a boundary, the entanglement entropy for an interval of length $\ell$ with one end on the boundary takes the form \cite{Calabrese:2004eu,Calabrese:2009qy}
\begin{align}
S_{\p} = \frac{c}{6} \ln \left(\frac{2 \ell}{a}\right) + \tilde c_1\,,
\end{align}
where $\tilde c_1$ is again a non-universal constant which in general is different than the $c_1$ appearing above.
The difference is well defined and yields the boundary entropy
\begin{align}
\label{entboundent}
\ln g = S_{\p}(\ell) - \frac{S(2\ell)}{2} = \tilde c_1 - \frac{\hat c_1}{2}\,.
\end{align}
This quantity was originally identified in \cite{Affleck:1991tk} as a contribution to the partition function which was independent of the size of the system.

At finite temperature the above formulas become
\begin{align}
\label{entTemp}
S =& \frac{c}{3} \ln \left( \frac{\beta}{\pi a} \sinh \frac{\pi\ell}{\beta} \right) + \hat c_1 \,, \cr
S_{\p} =& \frac{c}{6} \ln \left( \frac{\beta}{\pi a} \sinh \frac{2\pi\ell}{\beta} \right) + \tilde c_1\,,
\end{align}
with $\beta = T^{-1}$.
In particular, the presence of the boundary does not modify the thermal behavior of the entanglement entropy.  The above discussion naturally generalizes to the case of an $N$-junction.  We take $A$ to be the union of line segments of length $\ell$, where each segment has one endpoint located at the junction.  The entanglement entropy then takes the form
\begin{align}
\label{eenjunc}
S_{N} =& \sum_{i=1}^N \left[ \frac{c_i}{6} \ln \left( \frac{\beta}{\pi a} \sinh \frac{2\pi\ell}{\beta} \right)\right] + \tilde c_1
\end{align}
We may think of an $N$-junction as a boundary CFT, which is simply a product of the $N$-bulk CFTs with boundary conditions chosen so as to reproduce the $N$-junction theory.  If we take $N=2$ and $c_1 = c_2$, we recover the homogenous CFT results \eqref{entCFT} and \eqref{entTemp}, after taking into account that our segment has length $2 \ell$.

\begin{figure}[h!]
  \centering
    \includegraphics[width=0.7\textwidth]{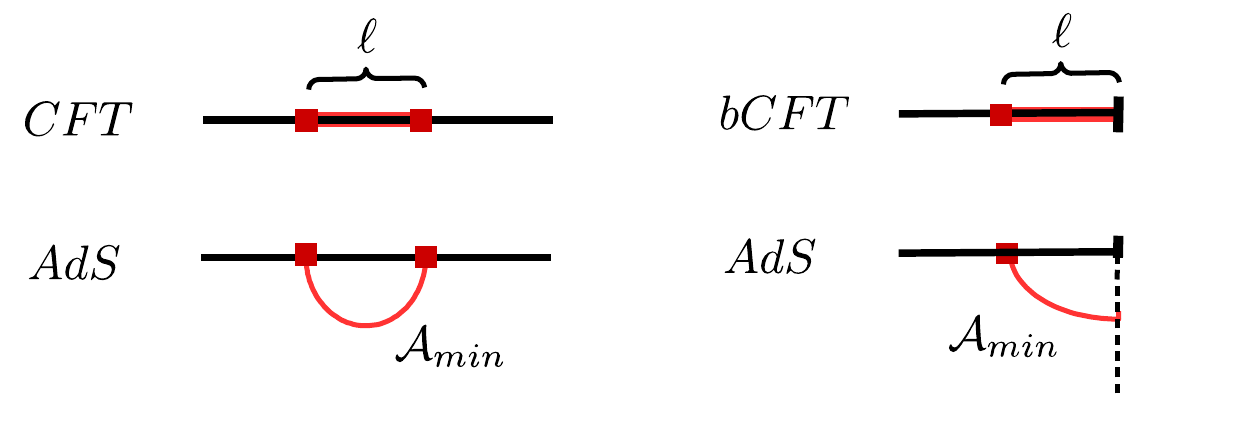}
    \caption{Schematic depiction of the entanglement line segment in the CFT (top) and the corresponding minimal curve in the dual $AdS_3$ space (bottom).  The left figure corresponds to the case of $2$-dimensional CFT, while the right corresponds to a $2$-dimensional boundary CFT (BCFT).  The dashed line on the right can correspond to an interior boundary, which runs into the bulk of $AdS$, or a smooth cap, where the geometry ends.}
    \label{fig2}
\end{figure}

The holographic prescription for computing entanglement entropy was originally proposed in \cite{Ryu:2006bv,Ryu:2006ef}.  First, the entangling surface is mapped onto the boundary of the dual geometry.  Then one computes the minimal area of a surface whose boundary is fixed to be the entangling surface.  The entanglement entropy is then given by
\begin{align}
S_N = \frac{{\cal A}_{min}}{4 G_N}\,,
\end{align}
where ${\cal A}_{min}$ is the minimal area and $G_N$ is Newton's constant.  For the case discussed above, the entangling surface consists of two points and we consider a line which begins and ends at these two points on the boundary of the $AdS$-space, as in figure \ref{fig2}.

In general the metric of our solution takes the form $ds^2 = f(y_a)^2 ds^2_{1,1} + ds^2_{y_a}$, where $ds^2_{y_a} = g_{ab} dy^a dy^b$ is the metric on the space spanned by the coordinates $y_a$ and $ds^2_{1,1} = -\rho^2 dt^2 + d\rho^2/\rho^2$ for the zero temperature case and $ds^2_{1,1} = -(\rho^2-1)dt^2 + d\rho^2/(\rho^2-1)$ for the finite temperature case.  We can choose coordinates so that the surface is parameterized by the coordinates $y_a$.  It then remains to find $\rho(y_a)$.
The area is given by
\begin{align}
{\cal A} = \int dw^a \sqrt{\det g_{ab}} \sqrt{1 + \frac{f^2}{\rho^2-1} g^{ab} \, \p_{a} \rho \, \p_{b} \rho} \,.
\end{align}
We have one plus a manifestly positive quantity appearing in the root.  The minimal value of the root is one and can be obtained by taking $\rho$ to be constant.  Thus the minimal area is given by the integral
\begin{align}
{\cal A}_{min} = \int dw^a \sqrt{\det g_{ab}} \,,
\end{align}
which is simply the volume of the space $ds^2_{w_a}$.
In general this is a divergent integral and must be regulated.  In order to define the boundary entropy as in \eqref{entboundent}, we should use the same regularization scheme for the two entanglement entropy computations.  The simplest way to achieve this is to use the map to Fefferman-Graham coordinates given in \eqref{CFTFGtrans} for zero temperature and in \eqref{flatFG} for finite temperature and impose the cutoff $z_c^{i} = \varepsilon / (L^i)^2$.

In general the Fefferman-Graham patch does not cover the entire manifold and contains only a single asymptotic $AdS_3$ region.  For an $N$-junction the manifold is decomposed into $N+1$ patches.  There is a central patch whose volume is finite and we denote by ${\cal A}^0_{min}$.  The remaining $N$ patches are Fefferman-Graham patches with a single asymptotic region.  An explicit example is provided by the solutions of \cite{Chiodaroli:2011nr} and we schematically depict the patching of a $4$-junction in figure \ref{fig1}.

In the $i$-th Fefferman-Graham patch, we may introduce coordinates $\{x,w^{\tilde a}\}$ so that the metric takes the form
\begin{align}
ds^2 = f^2 ds_{1,1}^2 + g_{xx} dx^2 + \tilde g_{\tilde a \tilde b} d\tilde w^{\tilde a} d\tilde w^{\tilde b}\,.
\end{align}
where $x^i_0 < x < \infty$, $x^i_0$ denotes the boundary of the Fefferman-Graham patch and the metric is asymptotically $AdS_3$ as $x \rightarrow \infty$.  The warp factors are finite for all values of $\tilde w^a$ and finite values of $x$.  For $x \rightarrow \infty$, we have $f \sim f^i_{-1} e^{x/L}$, $g_{xx} \sim 1$ and $\tilde g_{ab}$ is finite.  Using this behavior, the area of the minimal surface in the $i$-th patch takes the form
\begin{align}
\label{iarea}
{\cal A}^i_{min} =& \int d\tilde w^a \sqrt{\tilde g_{\tilde w}^0} \int_{x^i_0}^{x^i_c} dx \left(1 + \sum_{n=1}^\infty {\cal I}^i_n(x,\tilde w_a) e^{-n \frac{x}{L}} \right)
\cr
=& \int d\tilde w^a  \sqrt{\tilde g_{\tilde w}^0} \left[x^i_c - x^i_0 + {\cal F}^i(x^i_0,\tilde w_a) + {\cal O}(\varepsilon) \right] \,,
\end{align}
where we have introduced the functions ${\cal I}^i_n$ and ${\cal F}^i$.  The ${\cal I}^i_n$ arise as the sub-leading terms in the expansion of the integrand, $\sqrt{\det g_{ab}}$, in powers of $e^{-x}$, while ${\cal F}^i$ is defined in terms of the ${\cal I}^i_n$ by
\begin{align}
&\p_x {\cal F}^i(x,\tilde w_a) = \sum_{n=1}^\infty {\cal I}^{i}_n(x,\tilde w_a) e^{-n \frac{x}{L}}\,,&
&{\cal F}^i(\infty,w_a) = 0\,.&
\end{align}
As we will show below, the temperature dependence enters only through the cutoff $x^i_c$ and we have kept track of the explicit $x^i_c$ dependence up to terms which vanish in the $\varepsilon \rightarrow 0$ limit.

We now carefully introduce the cutoffs.  In the zero temperature theory, we take $r = \varepsilon$.  We also require the surface to intersect the boundary at $L_i y = \ell$, where $L_i$ is the radius of the $i$-th asymptotic $AdS_3$ region. Using the coordinate transformation \eqref{CFTFGtrans}, this gives the $x$-cutoff
\begin{align}
x^i_c =& - L^i \ln(\varepsilon) + L^i \ln \left( \frac{L^i \ell}{f^i_{-1}} \right) + {\cal O}(\varepsilon^2)\,.
\end{align}
The resulting entanglement entropy takes the form
\begin{align}
\label{gravEE}
S_N =& \sum_{i=1}^N \frac{c_i}{6} \, \ln\left(\frac{2 \ell}{\varepsilon}\right)
+ \tilde c_1
\,,\cr
c_i =& \frac{6 L^i}{4 G_N} \int d\tilde w^a  \sqrt{\tilde g_{\tilde w}^0} \,, \cr
\tilde c_1 =& \frac{{\cal A}^0_{min}}{4 G_N} + \frac{1}{4 G_N} \sum_{i=1}^N \int d\tilde w^a  \sqrt{\tilde g_{\tilde w}^0} \left[ L^i \ln \left( \frac{L_i}{2 f^i_{-1}} \right) - x^i_0 + {\cal F}^i(x^i_0,\tilde w_a)
\right]\,.
\end{align}
After identifying $a$ with the cutoff $\varepsilon$, this matches the zero temperature form of the CFT result given in \eqref{eenjunc}.

At finite temperature, we first choose coordinates so that our asymptotic solution matches the asymptotics of the BTZ black hole in \eqref{BTZ}. Using the same cutoff prescription as the zero temperature case, we impose the cutoff $z_c = L_i^2/\varepsilon$ along with the requirement that $L_i y' = \ell$ at the cutoff surface.  This leads to $r_c = z_0 \varepsilon / L_i^2 + {\cal O}(\varepsilon^2)$ and $y = \ell z_0/L_i^2$.  Using the coordinate transformation \eqref{flatFG}, this leads to the $x$-cutoff
\begin{align}
x_c^i =& -L_i \ln\left(\frac{2 \pi}{\beta} \varepsilon\right) + L_i \ln \left(  \frac{L_i \sinh\left(\frac{2 \pi \ell}{\beta}\right)}{f^i_{-1}} \right) + {\cal O}(\varepsilon^2)\,,
\end{align}
where we have used $z_0/L_i^2 = 2 \pi /\beta$.  Note that the $x_c^i$ are the only quantities which depend on $\beta$, since the decomposition of the manifold into Fefferman-Graham patches depends only on the behavior of $f$.
The resulting entanglement entropy is given by
\begin{align}
S_{N} =& \sum_{i=1}^N \left[ \frac{c_i}{6} \ln \left( \frac{\beta}{\pi a} \sinh \frac{2\pi\ell}{\beta} \right)\right] + \tilde c_1
\end{align}
where the $c_i$ and $\tilde c_1$ have the same definitions as in \eqref{gravEE}.  This exactly matches the finite temperature CFT result given in \eqref{eenjunc}.  Thus the temperature dependence corresponds to a thermal field theory as claimed.

\acknowledgments

J.E. is grateful to C. Bachas, E. D'Hoker, M. Gutperle, K. Jensen, A. O'Bannon, E. Stratos and T. Wrase for conversations and previous collaborations leading to this work.

\appendix

\section{Half-BPS string-junction solutions in six-dimensional supergravity}
\label{app:sixdcase}

In \cite{Chiodaroli:2011nr}, $SO(2,1) \times SO(3))$-invariant half-BPS solutions of $6$-dimensional $(0,4)$ supergravity with $m$ tensor multiplets were constructed.  The field content consists of a collection of scalars and $3$-forms in addition to the metric.  The scalars parameterize the coset $SO(5,m)/(SO(5)\times SO(m))$ and are organized into a collection of canonical frame and connection fields $P^{ir}$, $Q^{ij}$ and $S^{rs}$, with $i=1,..5$ and $r=6,...,m+5$.  The $3$-forms are organized into two sets, denoted by $H^i$ and $H^r$.  All equations of motion and Bianchi identities can be expressed in terms of these quantities.

Our goal in this appendix is to construct finite temperature generalizations of these solutions.  To do so, we take the metric to be given by a generalization of the one appearing in section 3 of \cite{Chiodaroli:2011nr}.  Namely, we replace the $AdS_2$ metric with an arbitrary 2-dimensional one so that
\begin{align}
ds^2 = f_1^2 ds^2_{1,1} + f_2^2 ds^2_{S^2} + ds^2_{\Sigma}\,,
\end{align}
where $ds^2_{1,1}$ does not depend on the $S^2$ or $\Sigma$.  Orthonormal frames are defined by
\begin{align}
&ds^2_{1,1} = \eta^{(2)}_{mn} \hat e^m \otimes \hat e^n&
&e^m = f_1 \hat e^m&
&m=0,1&\cr
&ds^2_{S^2} = \delta_{pq} \hat e^p \otimes \hat e^q&
&e^p = f_2 \hat e^p&
&p=2,3&\cr
&ds^2_{\Sigma} = \delta^{(2)}_{mn} e^a \otimes e^b&
&e^a = \rho \hat e^a&
&a=4,5\,.&
\end{align}
We also denote the frames collectively by the indices $M,N = 0,...,5$.  The only difference, as compared to \cite{Chiodaroli:2011nr}, is in the definition of $\hat e^m$.  Since the scalar field strength components are only supported along $\Sigma$, we take the scalars to be identical to the ones in \cite{Chiodaroli:2011nr}.  For the $3$-forms, we take
\begin{align}
&H^i = g^i_{a} e^{01a} + h_a^i e^{23a}\,,&
&H^r = \tilde g^r_a e^{01a} + \tilde h^r_a e^{23a}\,,&
\end{align}
where the coefficients are identical to the ones appearing in \cite{Chiodaroli:2011nr}.  The only difference is the volume form $e^{01a}$.

It will be useful to have the following expressions for exterior derivatives
\begin{align}
\label{extderivs}
& * d * e^{a} = - 2 *_{\Sigma} [(d \ln f_1) \wedge *_{\Sigma} e^{a}] - 2 *_{\Sigma} [(d \ln f_2) \wedge *_{\Sigma} e^{a}]
- *_{\Sigma} d *_{\Sigma} e^{a}\,,&\cr
&* d * e^{01a} = -e^{01} \wedge \{ 2 *_{\Sigma} [(d \ln f_2) \wedge *_{\Sigma} e^{a} ] + *_{\Sigma} d *_{\Sigma} e^{a} \} \,,& \cr
&* d * e^{23a} = -e^{23} \wedge \{ 2 *_{\Sigma} [(d \ln f_1) \wedge *_{\Sigma} e^{a} ] + *_{\Sigma} d *_{\Sigma} e^{a} \} \,,& \cr
&d e^{01a} = 2 (d \ln f_1) \wedge  e^{01a} - \omega^a{}_b e^{01b}\,,&
\end{align}
where $*_{\Sigma}$ denotes the Hodge dual with respect to the metric on $\Sigma$.  We note that the coefficients of these equations do not depend on the choice of metric for $ds^2_{1,1}$.

Now we show the equations of motion are automatically satisfied after taking $ds^2_{1,1}$ to be given by \eqref{Tfib}.  The Einstein equations are given by
\begin{align}
R_{MN} =  H^i_{MPQ} H^i_N {}^{PQ} + H^r_{MPQ} H^r_N {}^{PQ} + 2 P_M^{ir} P_N^{ir}
\end{align}
Since we do not change the scalar or 3-form components, meaning that the expressions for $H^i_{MNP}$, $H^r_{MNP}$, and $P_M^{ir}$ are identical to the ones appearing in \cite{Chiodaroli:2011nr}, the right hand side of the equation is unchanged after replacing $ds^2_{AdS_2}$ with an arbitrary metric $ds^2_{1,1}$.  A straightforward computation shows $R_{MN}$ takes the form
\begin{align}
&R_{mn} = \frac{R^{(1,1)}_{mn}}{f_1^2} + \eta_{mn} \left(-|D_a \ln f_1|^2 - 2(D^a \ln f_1)(D_a \ln f_2) - \frac{D^a D_a f_1}{f_1} + (\omega_a)^a{}_b  \, D^b \ln f_1 \right)  \,,&\no\\
&R_{pq} = \delta_{pq} \left(\frac{1}{f_2^2} - |D_a \ln f_2|^2 - 2(D^a \ln f_2)(D_a \ln f_1) - \frac{D^a D_a f_2}{f_2} + (\omega_a)^a{}_b  \, D^b \ln f_2  \right) \,,&\no\\
&R_{ab} = R^{(\Sigma)}_{ab} -2 \frac{D_b D_a f_1}{f_1} -2 \frac{D_b D_a f_2}{f_2} + (\omega_b)^a{}_b  \, D^b \ln f_1  + (\omega_b)^a{}_b  \, D^b \ln f_2\,,&
\end{align}
where the derivative $D_a$ is defined in terms of the partial derivative by $e^a D_a = dx^a \p_{x^a}$.  If we require the Ricci tensor for the $2$-dimensional metric $ds^2_{1,1}$ to satisfy $R^{(1,1)}_{mn} = - \eta_{mn}$, then the left hand side is also unchanged and we conclude that the Einstein equations are satisfied.

The field equations of the scalars and $3$-forms are given by
\begin{align}
&* d * P^{ir} = (Q^a)^{ij} P_a^{jr} + (S^a)^{rs} P_a^{is} + \frac{\sqrt{2}}{3} H^i {}^{MNP} H^r_{MNP}& \cr
&(* d * H^i)_{MN} = -(Q^P)^{ij} H_{PMN}^j + \sqrt{2} (P^P)^{ir} H_{PMN}^r \,,& \cr
&(* d * H^r)_{MN} = -(S^P)^{rs} H_{PMN}^j + \sqrt{2} (P^P)^{ir} H_{PMN}^i \,.&
\end{align}
Again, since we do not change the components, the right hand side of each equation is identical to the corresponding one appearing in \cite{Chiodaroli:2011nr}.  Using the first three lines appearing in \eqref{extderivs}, one can also see that the left hand side of each equation does not depend on the choice of metric $ds^2_{1,1}$.  As a result, we conclude the scalar and $3$-form equations of motion are satisfied.

Finally, we turn to the Bianchi identities.  The Bianchi identities for the scalar fields are clearly unmodified, since the scalar frame and connection $1$-forms have no support along $ds^2_{1,1}$.  The Bianchi identities for the $3$-forms are given by
\begin{align}
&dH^i - Q^{ij} \wedge H^j - \sqrt{2} P^{ir} \wedge H^r = 0\,,&
&dH^r - S^{rs} \wedge H^s - \sqrt{2} P^{ir} \wedge H^i = 0\,.&
\end{align}
Using the last line of \eqref{extderivs}, we see that again these equations will be satisfied, regardless of the choice of metric for $ds^2_{1,1}$.

\bibliographystyle{JHEP}
\bibliography{IBH3d3}

\providecommand{\href}[2]{#2}\begingroup\raggedright\begin{thebibliography}{10}

\bibitem{Chiodaroli:2009yw}
M.~Chiodaroli, M.~Gutperle, and D.~Krym, {\it {Half-BPS Solutions locally
  asymptotic to $AdS_3 \times S^3$ and interface conformal field theories}},
  {\em JHEP} {\bf 1002} (2010) 066,
  [\href{http://xxx.lanl.gov/abs/0910.0466}{{\tt arXiv:0910.0466}}].

\bibitem{Chiodaroli:2009xh}
M.~Chiodaroli, E.~D'Hoker, and M.~Gutperle, {\it {Open Worldsheets for
  Holographic Interfaces}},  {\em JHEP} {\bf 1003} (2010) 060,
  [\href{http://xxx.lanl.gov/abs/0912.4679}{{\tt arXiv:0912.4679}}].

\bibitem{Chiodaroli:2011nr}
M.~Chiodaroli, E.~D'Hoker, Y.~Guo, and M.~Gutperle, {\it {Exact half-BPS
  string-junction solutions in six-dimensional supergravity}},  {\em JHEP} {\bf
  1112} (2011) 086, [\href{http://xxx.lanl.gov/abs/1107.1722}{{\tt
  arXiv:1107.1722}}].

\bibitem{Chiodaroli:2012vc}
M.~Chiodaroli, E.~D'Hoker, and M.~Gutperle, {\it {Holographic duals of Boundary
  CFTs}},  {\em JHEP} {\bf 1207} (2012) 177,
  [\href{http://xxx.lanl.gov/abs/1205.5303}{{\tt arXiv:1205.5303}}].

\bibitem{Chiodaroli:2011fn}
M.~Chiodaroli, E.~D'Hoker, and M.~Gutperle, {\it {Simple Holographic Duals to
  Boundary CFTs}},  {\em JHEP} {\bf 1202} (2012) 005,
  [\href{http://xxx.lanl.gov/abs/1111.6912}{{\tt arXiv:1111.6912}}].

\bibitem{Kondo01071964}
J.~Kondo, {\it Resistance minimum in dilute magnetic alloys},  {\em Progress of
  Theoretical Physics} {\bf 32} (1964), no.~1 37--49,
  [\href{http://xxx.lanl.gov/abs/http://ptp.oxfordjournals.org/content/32/1/37.full.pdf+html}{{\tt
  http://ptp.oxfordjournals.org/content/32/1/37.full.pdf+html}}].

\bibitem{PhysRevLett.12.442}
M.~Strongin, A.~Paskin, D.~G. Schweitzer, O.~F. Kammerer, and P.~P. Craig, {\it
  Surface superconductivity in type i and type ii superconductors},  {\em Phys.
  Rev. Lett.} {\bf 12} (Apr, 1964) 442--444.

\bibitem{PhysRevLett.49.405}
D.~J. Thouless, M.~Kohmoto, M.~P. Nightingale, and M.~den Nijs, {\it Quantized
  hall conductance in a two-dimensional periodic potential},  {\em Phys. Rev.
  Lett.} {\bf 49} (Aug, 1982) 405--408.

\bibitem{2009AIPC.1134...22K}
A.~{Kitaev}, {\it {Periodic table for topological insulators and
  superconductors}},  in {\em American Institute of Physics Conference Series}
  (V.~{Lebedev} and M.~{Feigel'Man}, eds.), vol.~1134 of {\em American
  Institute of Physics Conference Series}, pp.~22--30, May, 2009.
\newblock \href{http://xxx.lanl.gov/abs/0901.2686}{{\tt arXiv:0901.2686}}.

\bibitem{RevModPhys.82.3045}
M.~Z. Hasan and C.~L. Kane, {\it Colloquium},  {\em Rev. Mod. Phys.} {\bf 82}
  (Nov, 2010) 3045--3067.

\bibitem{Bak:2011ga}
D.~Bak, M.~Gutperle, and R.~A. Janik, {\it {Janus Black Holes}},  {\em JHEP}
  {\bf 1110} (2011) 056, [\href{http://xxx.lanl.gov/abs/1109.2736}{{\tt
  arXiv:1109.2736}}].

\bibitem{Bachas:2011xa}
C.~Bachas and J.~Estes, {\it {Spin-2 spectrum of defect theories}},  {\em JHEP}
  {\bf 1106} (2011) 005, [\href{http://xxx.lanl.gov/abs/1103.2800}{{\tt
  arXiv:1103.2800}}].

\bibitem{Calabrese:2004eu}
P.~Calabrese and J.~L. Cardy, {\it {Entanglement entropy and quantum field
  theory}},  {\em J.Stat.Mech.} {\bf 0406} (2004) P06002,
  [\href{http://xxx.lanl.gov/abs/hep-th/0405152}{{\tt hep-th/0405152}}].

\bibitem{Calabrese:2009qy}
P.~Calabrese and J.~Cardy, {\it {Entanglement entropy and conformal field
  theory}},  {\em J.Phys.} {\bf A42} (2009) 504005,
  [\href{http://xxx.lanl.gov/abs/0905.4013}{{\tt arXiv:0905.4013}}].

\bibitem{Bak:2003jk}
D.~Bak, M.~Gutperle, and S.~Hirano, {\it {A Dilatonic deformation of AdS(5) and
  its field theory dual}},  {\em JHEP} {\bf 0305} (2003) 072,
  [\href{http://xxx.lanl.gov/abs/hep-th/0304129}{{\tt hep-th/0304129}}].

\bibitem{Korovin:2013gha}
Y.~Korovin, {\it {First order formalism for the holographic duals of defect
  CFTs}},  {\em JHEP} {\bf 1404} (2014) 152,
  [\href{http://xxx.lanl.gov/abs/1312.0089}{{\tt arXiv:1312.0089}}].

\bibitem{MR837196}
C.~Fefferman and C.~R. Graham, {\it Conformal invariants},  {\em Ast\'erisque}
  (1985), no.~Numero Hors Serie 95--116. The mathematical heritage of {\'E}lie
  Cartan (Lyon, 1984).

\bibitem{deHaro:2000xn}
S.~de~Haro, S.~N. Solodukhin, and K.~Skenderis, {\it {Holographic
  reconstruction of space-time and renormalization in the AdS / CFT
  correspondence}},  {\em Commun.Math.Phys.} {\bf 217} (2001) 595--622,
  [\href{http://xxx.lanl.gov/abs/hep-th/0002230}{{\tt hep-th/0002230}}].

\bibitem{Estes:2014hka}
J.~Estes, K.~Jensen, A.~O'Bannon, E.~Tsatis, and T.~Wrase, {\it {On Holographic
  Defect Entropy}},  {\em JHEP} {\bf 1405} (2014) 084,
  [\href{http://xxx.lanl.gov/abs/1403.6475}{{\tt arXiv:1403.6475}}].

\bibitem{Takayanagi:2011zk}
T.~Takayanagi, {\it {Holographic Dual of BCFT}},  {\em Phys.Rev.Lett.} {\bf
  107} (2011) 101602, [\href{http://xxx.lanl.gov/abs/1105.5165}{{\tt
  arXiv:1105.5165}}].

\bibitem{Erdmenger:2014xya}
J.~Erdmenger, M.~Flory, and M.-N. Newrzella, {\it {Bending branes for DCFT in
  two dimensions}},  {\em JHEP} {\bf 1501} (2015) 058,
  [\href{http://xxx.lanl.gov/abs/1410.7811}{{\tt arXiv:1410.7811}}].

\bibitem{Holzhey:1994we}
C.~Holzhey, F.~Larsen, and F.~Wilczek, {\it {Geometric and renormalized entropy
  in conformal field theory}},  {\em Nucl.Phys.} {\bf B424} (1994) 443--467,
  [\href{http://xxx.lanl.gov/abs/hep-th/9403108}{{\tt hep-th/9403108}}].

\bibitem{Affleck:1991tk}
I.~Affleck and A.~W. Ludwig, {\it {Universal noninteger 'ground state
  degeneracy' in critical quantum systems}},  {\em Phys.Rev.Lett.} {\bf 67}
  (1991) 161--164.

\bibitem{Ryu:2006bv}
S.~Ryu and T.~Takayanagi, {\it {Holographic derivation of entanglement entropy
  from AdS/CFT}},  {\em Phys.Rev.Lett.} {\bf 96} (2006) 181602,
  [\href{http://xxx.lanl.gov/abs/hep-th/0603001}{{\tt hep-th/0603001}}].

\bibitem{Ryu:2006ef}
S.~Ryu and T.~Takayanagi, {\it {Aspects of Holographic Entanglement Entropy}},
  {\em JHEP} {\bf 0608} (2006) 045,
  [\href{http://xxx.lanl.gov/abs/hep-th/0605073}{{\tt hep-th/0605073}}].

\end{thebibliography}\endgroup

\end{document}